\documentstyle[11pt,newpasp,twoside,epsf]{article} 
\def\edcomment#1{\iffalse\marginpar{\raggedright\sl#1\/}\else\relax\fi} 
\reversemarginpar 

\begin{document} 
\title{The relative ages of LMC old clusters, and the case of NGC~1841}

\author{Ivo Saviane, Alfred Rosenberg, Giampaolo Piotto, \& Antonio Aparicio}
\affil{ESO Chile, IAC, UniPD, IAC}

\begin{abstract} 
Using archival HST/WFPC2 imaging of $7$ LMC globular clusters, and
following the methods outlined in 
our previous study,
we have
reached the tightest constrain so far on their age dispersion, which
cannot be greater than $\approx 0.5$~Gyr. We also confirm earlier
results that their average age is comparable to that of the metal-poor
Galactic globulars. Evidence is also provided that NGC~1841 is younger
than the rest of LMC globulars.
\end{abstract}

\section{Introduction}

The LMC is the closest disk galaxy (at \( \sim 50 \) kpc) with a
large ensamble of star clusters. At least \( 11 \) of them can be
considered
\emph{ bona fide} counterparts (e.g. Olszewski et al. 1996) 
of Galactic globular clusters (GGC).
Their absolute ages are comparable to those of
the Milky Way  metal-poor ones (Brocato et al. 1996; Olsen et
al. 1998; Johnson et al. 1999), however, 
their kinematics 
is strikingly different, since
they show a definite disk-like rotation of
$\approx 80~\rm km~s^{-1}$ (Freeman
et al.
1983).
It is of great interest to understand the reasons of the simultaneous
formation of the two systems in such dynamically different
protogalaxies.  A key input to this problem is the age dispersion of LMC
globulars.

\section{The Relative ages of LMC old clusters, and the case of NGC~1841}

\begin{figure}  
\plotone{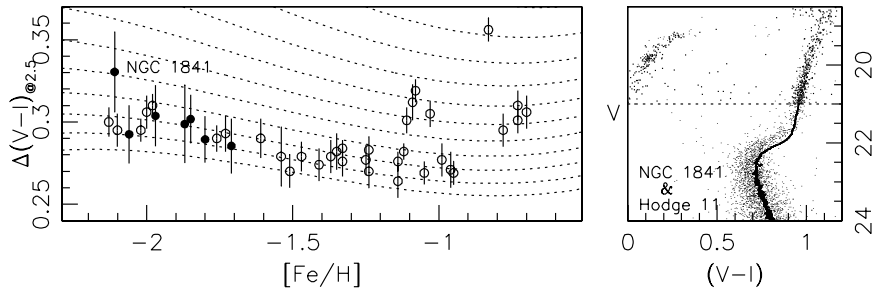} 
\caption{Left panel: the run of our  age-sensitive parameter
vs. metallicity, for
both Galactic globulars (open circles) and LMC ones (filled circles).
The dotted lines represent the Vandenberg et al. (2000) models, in $1$
Gyr steps (the bottommost line corresponding to $18$~Gyr). 
Right panel: 
The 
CMD of Hodge~11  has been
shifted in color and magnitude so that its RGB and HB overlap those of
NGC~1841. The good 
match
is espected from
the small difference in metallicity (Palmieri et al. 2002).  Fainter
than $V=21$ just a fiducial is shown for Hodge~11.
\label{fig:twopanels}}
\end{figure}

The CMDs are based on archival HST/WFPC2 data (GO 5897). Data reduction
and calibration will be presented in a forthcoming paper.
In Fig.~1 (left panel)
the trend of \( \delta (V-I)_{@2.5} \) vs. [Fe/H]
is compared to a set of theoretical isochrones (see R99 for details), 
and 
the six LMC clusters fall inside the $\approx 2$~Gyr
strip defined by metal-poor GGCs. We thus confirm that most LMC clusters
are coeval with the Milky Way globulars, and more importantly, for the
first time, we can constrain the age dispersion of the LMC flattened halo,
which results to be as low as
$\sigma_t \approx 0.5$~Gyr.
According to Fig.~1 (left panel), 
NGC~1841 is $\sim 2~$~Gyr younger than the mean Galactic/LMC sample, a
fact that is confirmed by a direct comparison of its CMD to that of
Hodge~11 (right panel of Fig.~1). Indeed, one can see that the young age
is revealed by the HB-TO difference as well,
since the
NGC~1841 turnoff is brighter than that of Hodge~11.

\section{Conclusions}

We have found that both the mean age and age dispersion of the LMC GCs
is comparable to that of the metal-poor Galactic GCs, and clear evidence
has been obtained in favor of a younger age for NGC~1841.  Since it is
the most metal-poor LMC cluster,
it means that the interstellar medium (ISM) where it formed, 
has had a 
lower degree
of chemical evolution.
NGC~1841 could have been formed in a relatively isolated
fragment of the proto-LMC, or it could have been part of an independent
system now disrupted. Further support to the latter hypothesis could be
the fact that NGC~1841 is the farthest cluster from the LMC center
($\sim 10$~kpc)
and  that its radial velocity is incompatible with 
the  rotation 
of
the old LMC halo (Freeman et al. 1983).
The fact that LMC globulars formed in a rotating protogalaxy
with perfect synchronization with the 
GGCs,
seems to require
a joint formation scenario, possibly 
induced 
by the Milky Way early SF (e.g. Taniguchi et al. 1999).

\end{document}